# Ultrafast demagnetization of iron induced by optical vs terahertz pulses

A.L. Chekhov[1,2], Y. Behovits[1,2], J.J.F. Heitz[1,2], C. Denker[3], D.A. Reiss[1], M. Wolf[2], M. Weinelt[1], P.W. Brouwer[1], M. Münzenberg[3] and T. Kampfrath[1,2]

1. Fachbereich Physik, Freie Universität Berlin, Arnimallee 14, 14195 Berlin, Germany
2. Fritz-Haber-Institut der Max-Planck-Gesellschaft, Faradayweg 4-6, 14195 Berlin, Germany
3. Institut für Physik, Universität Greifswald, Felix-Hausdorff-Straße 6, 17489 Greifswald, Germany

We study ultrafast magnetization quenching of ferromagnetic iron following excitation by an optical vs a terahertz pump pulse. While the optical pump (photon energy of 3.1 eV) induces a strongly nonthermal electron distribution, terahertz excitation (~4 meV) results in a quasi-thermal perturbation of the electron population. The pump-induced spin and electron dynamics are interrogated by the magneto-optic Kerr effect (MOKE). A deconvolution procedure allows us to push the time resolution down to 130 fs, even though the driving terahertz pulse is more than 0.5 ps long. Remarkably, the MOKE signals exhibit an almost identical time evolution for both optical and terahertz pump pulses, despite the three orders of magnitude different number of excited electrons. We are able to quantitatively explain our results using a model based on quasi-elastic spin-flip scattering. It shows that in the small-perturbation limit, the rate of demagnetization of a metallic ferromagnet is proportional to the excess energy of the electrons, independent of the precise shape of their distribution. Our results reveal that the dynamics of ultrafast demagnetization and of the closely related terahertz spin transport do not depend on the pump photon energy.

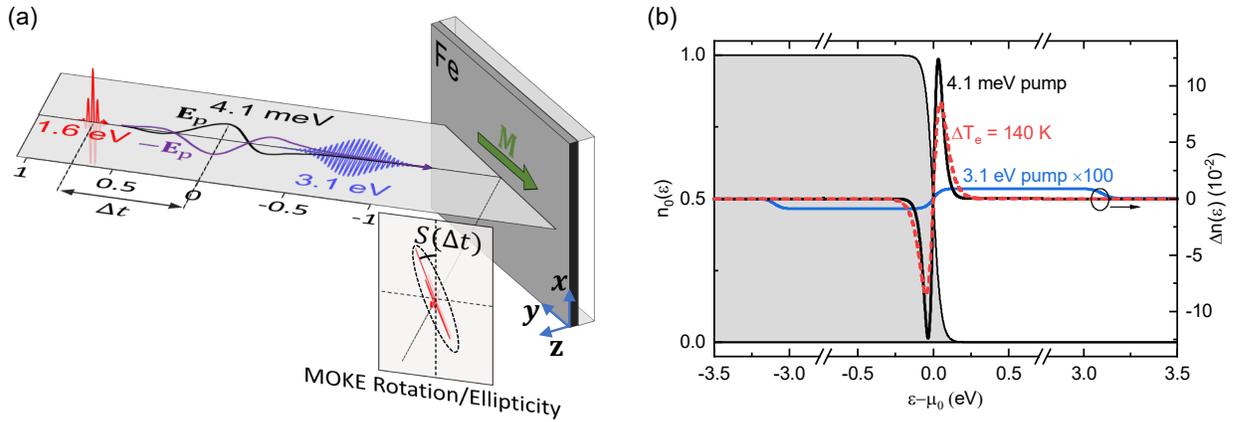

Figure 1. (a) Schematic of the experiment. A pump pulse at optical (photon energy of 3.1 eV) or terahertz (4.1 meV) frequency with transient electric field $E_p(t)$ excites a ferromagnetic Fe thin film with in-plane magnetization $M_0$ along the $y$ axis. The subsequent sample evolution is monitored by a time-delayed probe pulse (1.6 eV), whose polarization state is detected after reflection from the sample. (b) Calculated variations $\Delta n(\epsilon)$ of the electron occupation number vs electron energy $\epsilon$ after deposition of 0.01 mJ/cm² by photons with 4.1 meV (black curve) and 3.1 eV energy (blue curve). We assumed constant electronic density of states and transition matrix elements (see Supplemental Materials). For the THz pump, $\Delta n(\epsilon)$ can be well fitted by the difference of two Fermi-Dirac distributions with temperatures $T_0 + 140$ K and $T_0 = 300$ K and the same chemical potential $\mu_0$ (red curve). The shaded area shows the unperturbed Fermi-Dirac distribution $n_0(\epsilon)$ with temperature $T_e = T_0 = 300$ K and chemical potential $\mu_0$.

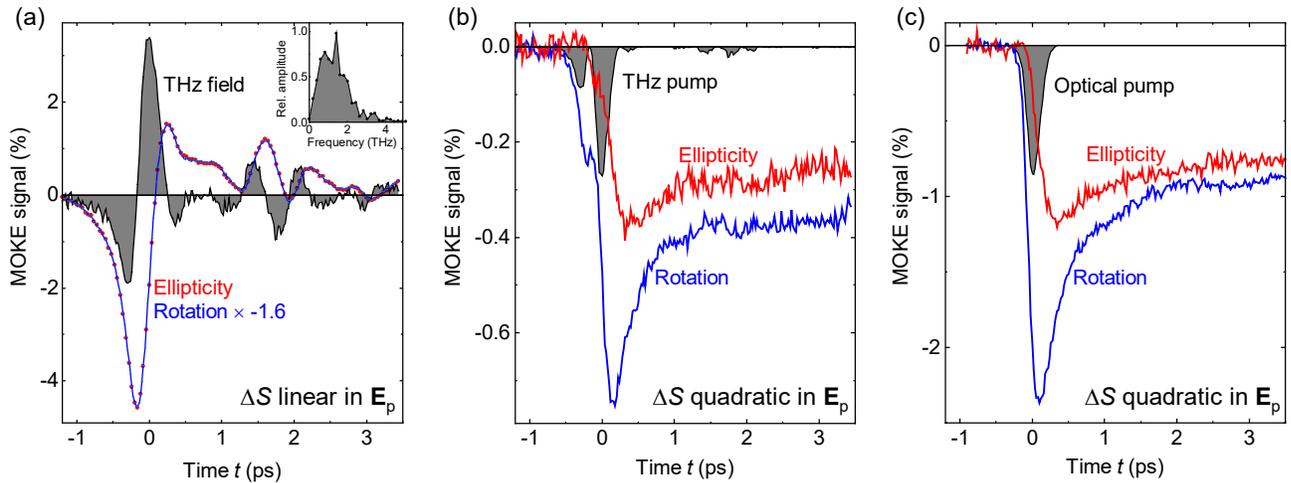

Figure 2. Typical traces of pump-induced MOKE rotation and ellipticity $\Delta S(t)$ odd in $M_0$. (a) Rotation (blue curve, scaled) and ellipticity (red) signals linear in the field $E_p$ of the driving THz pump pulse. Assuming this contribution arises from Zeeman torque, its derivative is proportional to the THz magnetic and electric field (black shaded curve). The inset shows the amplitude spectrum of the THz pump field $E_p$. (b) THz-pump-induced MOKE rotation (blue) and ellipticity (red) quadratic in $E_p$. (c) Same as panel (b), but for the optical pump pulse. Shaded areas (grey) indicate the corresponding pump intensity profiles $E_p^2(t)$.

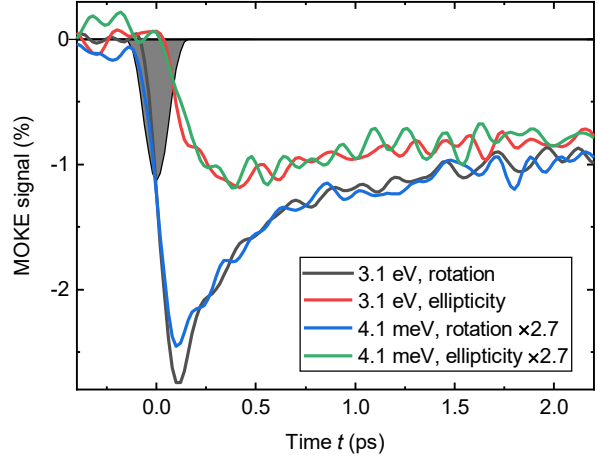

Figure 3. Pump-induced variations of MOKE rotation and ellipticity odd in the sample magnetization $M_0$ and quadratic in the pump field $E_p$. All signals are deconvoluted with respect to the intensity profile $E_p^2(t)$ of optical (3.1 eV) and terahertz (4.1 meV) pump pulses. They refer to the same fictitious pump profile $E_{fic}^2(t)$ as shown by the shaded black curve. Signals for the terahertz-pump data are scaled to match the optical one. The scaling factor of 2.7 corresponds to the ratio of the absorbed optical and THz pump fluences (see Supplemental Materials).

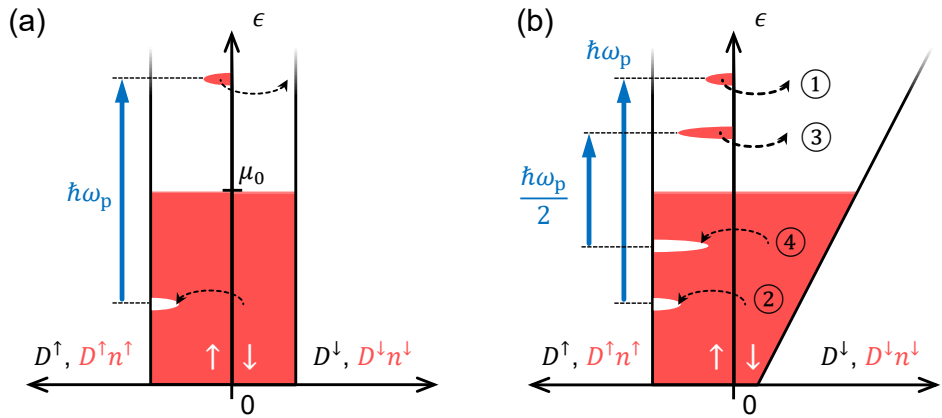

Figure 4. Two model scenarios of optically induced magnetization quenching. (a) The electronic densities of states $D^\uparrow(\epsilon)$ and $D^\downarrow(\epsilon)$ (black lines) and transition-matrix elements are independent of electron energy $\epsilon$. The total number $n^\sigma(\epsilon)D^\sigma(\epsilon)$ of electrons with spin $\sigma = \uparrow$ or $\downarrow$ and energy $\epsilon$ is indicated by the red area. The pump pulse (photon energy $\hbar\omega_p$, blue arrow) generates electrons and holes above and below the Fermi level $\mu_0$, thereby modifying the number of spin-up electrons $D^\uparrow n^\uparrow$ at a given energy $\epsilon$. Because $D^\uparrow$ and $D^\downarrow$ do not depend on $\epsilon$, the net number of spin-flip scattering events is zero, leaving the magnetization unchanged. (b) The spin-down density of states $D^\downarrow$ increases linearly with $\epsilon$. Consequently, more scattering events for the spin-up electrons (①) than for the spin-up holes (②) are possible, resulting in magnetization quenching. When the photon energy is reduced to, for instance, $\hbar\omega_p/2$, the difference in the number of final spin-down states for a spin-up electron (③) minus the analogous number for a spin-up hole (④) decreases by a factor of 2. This effect is, however, compensated by the number of pump-excited spin-up electrons, which has increased by a factor of 2. The resulting demagnetization rate is the same as for photon energy $\hbar\omega_p$, provided the deposited pump energy is the same.

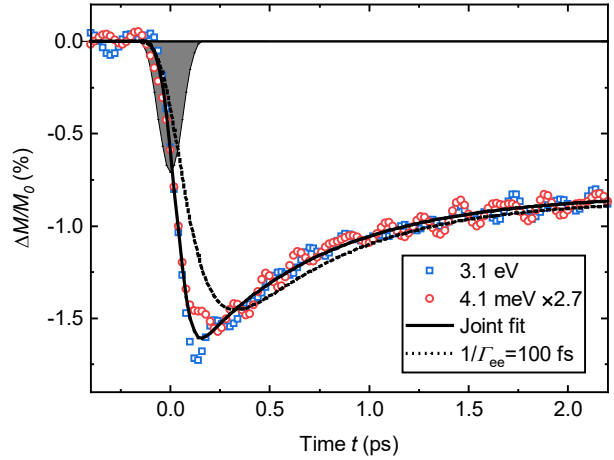

Figure 5. Extracted magnetization dynamics $\Delta M(t)/M_0$ for optical (3.1 eV, blue squares) and terahertz (4.1 meV, red circles) excitation with a temporal profile indicated by the grey area. The data is inferred from the transients shown in Fig.3. The black solid curve is a fit as described in the text. The dotted curve is a result of a fit for THz excitation and subsequent convolution with a step-like exponential (decay time 100 fs) mimicking a situation where electron-electron thermalization after optical excitation is relevant.

**Introduction**

To push writing of magnetic information to ultrafast time scales, it is essential to understand the response of magnetically ordered solids to a quasi-instantaneous perturbation [1,2]. A model experiment is the excitation of a ferromagnet by a femtosecond laser pulse [Fig. 1(a)] [3–5]. In ferromagnetic metals like Fe, Co and Ni, the resulting ultrafast demagnetization (UDM) is known to proceed on a time scale of 100 fs and yields insights into fundamental material parameters such as the electron-spin equilibration time. In terms of applications, UDM is an important process in all-optical magnetization switching [2]. It is driven by the same force as ultrafast spin transport [6,7] and, thus, of relevance for applications such as generation of spin torque [8] and terahertz electromagnetic pulses [9–11].

An open question regarding UDM is how strongly the magnetization dynamics depends on the nature of the pump-induced electron distribution. In thermal equilibrium, the electron distribution is Fermi-Dirac-like [see Fig. 1(b)]. It becomes strongly nonthermal [12–17] by excitation with an optical pulse, as schematically shown by [Fig. 1(b)] for a pump photon energy of 3.1 eV. On a time scale of ~100 fs, electron scattering (in particular with other electrons) again leads to a Fermi-Dirac (thermal) distribution with elevated electron temperature $T_e$ [18–20].

Several works emphasize that the initial nonthermal electron distribution can lead to an enhanced quenching rate of the magnetization [14,21,22], for instance through Bloch states with large spin-orbit coupling [14], so-called hot spots. Other works, however, argue that the nonthermal character of the initial electron distribution is rather negligible in terms of UDM [5,23,24].

To study the impact of the precise shape of the initial pump-induced electron distribution, we propose to use a pump photon energy either much larger (3.1 eV) or smaller (4.1 meV) than the characteristic thermal energy of $k_B T_0 = 25$ meV, where $k_B$ is the Boltzmann constant and $T_0 = 300$ K is the equilibrium sample temperature under ambient conditions [Fig. 1(b)]. Straightforward model calculations (see Supplemental Material) show that this approach allows us to tune the character of the electron distribution right after excitation from highly nonthermal [blue curve in Fig. 1(b)] to Fermi-Dirac-like [black curve in Fig. 1(b)]. It is remarkable that initially and for the same deposited energy, the THz pump yields three orders of magnitude more excited electrons than the optical pump. These two distributions imply a very different phase space for subsequent electron scattering and, thus, potentially different magnetization dynamics.

Previous works have already shown that UDM occurs at pump photon energies below 1.5 eV, from the infrared [25] down to the THz range [26–28]. As the THz pulses have a center frequency of typically 1 THz, it is, however, challenging to push the time resolution to the characteristic time constants of electron-spin (~100 fs) [4,7], electron-electron (~100 fs) [20,29] and electron-phonon equilibration (~0.5 ps) [4].

In this letter, we measure the ultrafast magnetization dynamics of the model ferromagnet Fe following ultrashort THz (4.1 meV) vs optical (3.1 eV) excitation. To directly compare the two data sets, we stay in the linear-with-fluence regime and apply a deconvolution procedure, thereby pushing the time resolution down to 130 fs, even though the driving THz pulse has a duration of more than 0.5 ps. Surprisingly, the spin dynamics are found to proceed almost identically for both THz (thermal) and optical (nonthermal) sample excitation. Therefore, the evolution of UDM is predominantly determined by just the amount of deposited pump energy, rather than the precise shape of the pump-induced electron distribution. We explain this behavior by an analytical nonthermal model based on quasi-elastic spin-flip scattering.

**Experimental details.** A schematic of our pump-probe experiment is displayed in Fig. 1(a), and details are shown in Fig. S1(a). A ferromagnetic Fe thin film is excited either by (i) an optical or (ii) a THz pump pulse under an angle of incidence of 45°. The resulting evolution of the electron spin and orbital degrees of freedom is monitored by a time-delayed near-infrared probe pulse, whose polarization state after reflection off the sample is measured.

As sample, we choose a ferromagnetic Fe thin film on a glass substrate and capped with 3 nm $SiO_x$. The Fe-layer thickness of 4 nm is significantly smaller than the attenuation length of the pump (~80 nm at 1 THz, ~10 nm at 3.1 eV) and the probe (~20 nm) [30]. Together with the insulating substrate, this condition prevents transport due to field gradients along the sample normal. The sample is characterized with THz transmission spectroscopy (Fig. S4) and yields an electron-velocity relaxation time of $\approx 10$ fs. When the pump is off, the sample magnetization $\boldsymbol{M}_0 = M_0 \boldsymbol{u}_y$ is in the plane and parallel to the $y$ axis with unit vector $\boldsymbol{u}_y$ [Fig. 1(a)].

In our setup, all laser pulses are obtained from an amplified Ti:sapphire laser system. The THz pump pulses (photon energy 4.1 meV, duration 500 fs, pulse energy 1 µJ, repetition rate 1 kHz) are generated by optical rectification of amplified laser pulses (1.55 eV, 50 fs, 4 mJ, 1 kHz) in a $LiNbO_3$ crystal [31,32]. The transient electric field and amplitude spectrum of the THz pulse are, respectively, shown in [Fig. 2(a)]. The optical pump pulses (3.1 eV, 200 fs, 140 nJ, 1 kHz) are obtained by frequency-doubling of the amplified pulses, and their intensity profile is shown in Fig. 2(c) (gray area). Because the refractive index of Fe at ~1 THz is large (~250) [30], the THz pump field inside the Fe film is in-plane to very good approximation.

As probe, we use pulses (1.6 eV, 40 fs, 1 nJ, 80 MHz) from the seed oscillator of the laser system (Fig.S1a) [33,34]. Changes in polarization rotation and ellipticity of the probe upon reflection from the sample are measured by a balanced detection scheme as a function of the delay $t$ since sample excitation by the pump pulse at $t = 0$.

**Expected signals.** The detected signal $S(t)$, which equals either the rotation ($S^{\mathrm{rot}}$) or ellipticity ($S^{\mathrm{ell}}$) of the probe polarization, is up to linear order in the sample magnetization $\boldsymbol{M}$ given by [35–37] $S = \boldsymbol{a} \cdot \boldsymbol{M} + b$. The first term is the magneto-optic Kerr effect (MOKE), where the vector $\boldsymbol{a} = \partial S/\partial \boldsymbol{M}$ quantifies how strongly the magnetization affects the linear optical properties of the sample. A nonmagnetic contribution $b$ appears due to a difference between Fresnel reflection coefficients for s and p polarization.

The pump pulse modifies the quantities $\boldsymbol{a}_0$, $\boldsymbol{M}_0 = M_0 \boldsymbol{u}_y$ and $b_0$ of the unperturbed sample by $\Delta \boldsymbol{a}$, $\Delta \boldsymbol{M}$ and $\Delta b$, respectively. We, thus, measure the pump-induced signal $\Delta S = \Delta(\boldsymbol{a} \cdot \boldsymbol{M} + b)$ as a function of $t$, which equals

$$\Delta S = \Delta(\boldsymbol{a} \cdot \boldsymbol{M} + b) = \boldsymbol{a}_0 \cdot \Delta \boldsymbol{M} + M_0 \Delta a_y + \Delta b \qquad (1)$$

up to first order in the pump-induced changes. Note that Eq. (1) applies to both MOKE rotation $S^{\mathrm{rot}}(t)$ and ellipticity $S^{\mathrm{ell}}(t)$. The coupling constants are, however, expected to exhibit disparate dynamics [35–38], $\Delta \boldsymbol{a}^{\mathrm{rot}}(t) \neq \Delta \boldsymbol{a}^{\mathrm{ell}}(t)$.

We expect two contributions to $\Delta \boldsymbol{M}$: (i) a component $\Delta \boldsymbol{M}|_{\mathrm{ZT}} = \Delta M_z \boldsymbol{u}_z$ perpendicular to the sample plane, which arises from Zeeman-type torque by the magnetic component $\boldsymbol{B}_{\mathrm{p}}(t)$ of the pump field [27,39], and (ii) a contribution $\Delta \boldsymbol{M}|_{\mathrm{UDM}} = \Delta M_y \boldsymbol{u}_y$ along the direction of the unperturbed $\boldsymbol{M}_0$ due to UDM [26]. To lowest order, the contributions (i) and (ii) scale linearly and quadratically in the pump field $\boldsymbol{E}_{\mathrm{p}}$, respectively. While component (ii) occurs for both THz and optical pump, component (i) oscillates rapidly with the optical pump field and, thus, cannot be time-resolved with our probe.

To suppress effects of third and higher order in $\boldsymbol{E}_{\mathrm{p}}$, we choose the absorbed pump fluence below 0.1 mJ cm$^{-2}$ (see Supplemental Material). To separate the various signal contributions, we alternate the sign of $\boldsymbol{M}_0$ and THz pump field $\boldsymbol{E}_{\mathrm{p}}$. First, measuring for $\pm \boldsymbol{M}_0$ allows us to separate $\Delta b$ and $\Delta(\boldsymbol{a} \cdot \boldsymbol{M})$ (see Eq. (1)). Second, alternating the sign of the THz field $\boldsymbol{E}_{\mathrm{p}}$ allows us to separate the contributions related to magnetization quenching (linear in $\boldsymbol{M}_0$, quadratic in $\boldsymbol{E}_{\mathrm{p}}$) and Zeeman torque (linear in both $\boldsymbol{M}_0$ and $\boldsymbol{E}_{\mathrm{p}}$, see Supplemental Material). Hysteresis-loop measurements of the unexcited sample yield the value of $a_{0y} M_0$, which is used to normalize the pump-probe traces (see Supplemental Material and Eq. (9)).

**Pump-probe signals.** Figure 2 displays MOKE signals obtained following THz [Figs. 2(a,b)] and optical pumping [Fig. 2(c)]. The full set of the underlying raw data is shown in Fig. S2. All signals depend linearly or quadratically on the pump field amplitude (Fig. S3).

Figure 2(a) shows signals $\Delta S$ following THz excitation, odd in $M_0$ and linear in $E_\mathrm{p}$. Note that rotation and ellipticity signal exhibit the same dynamics, which indicates that we observe exclusively true magnetization dynamics here ($\Delta a_y = 0$). Indeed, in electric-dipole approximation, a pump-induced change $\Delta a$ odd in $E_\mathrm{p}$ is not allowed in the bulk of our nominally inversion-symmetric Fe thin film. A nonvanishing magnetization change, on the other hand, is only possible in magnetic-dipole approximation and exclusively affects $M_z$. This fact is fully consistent with Zeeman torque as the microscopic excitation mechanism, which obeys $\Delta \dot{M}\big|_\mathrm{ZT} \propto B_\mathrm{p}(t) \times M_0$ (see Supplemental Material). Therefore, by taking the time derivative of the MOKE signals of Fig. 2(a), we consistently obtain a waveform which is proportional to the THz magnetic and electric field inside the Fe film (see Fig. S2b).

Figure 2(b) shows MOKE signals following THz excitation, odd in $M$ and quadratic in $E_\mathrm{p}$. The signals rise over the duration of the pump pulse and decays subsequently. The double-peak structure of the THz-driven dynamics coincides with the double peak in the THz intensity profile [shaded area in Fig. 2(b)]. Figure 2(c) displays corresponding signals $\Delta S$ following optical excitation. Analogous to the THz case, the $\Delta S$ rise with arrival of the pump and decay subsequently.

Note that the rotation and ellipticity signals in Figs. 2(b) and 2(c) exhibit substantially different dynamics. Therefore, the signal $\Delta S$ quadratic in the pump field $E_\mathrm{p}$ contains pump-induced variations $\Delta a$ of the magneto-optic constants (Eq. (1)) and, thus, cannot be interpreted as pure magnetization dynamics. This point is, however, no issue here because we are primarily interested in a comparison of THz and optical pumping. Further below, we will approximately extract $\Delta M(t)$ from our data.

**Deconvoluting the pump profile.** We aim to compare the signals following optical and THz excitation quadratic in the pump field $E_\mathrm{p}$. This goal is not straightforward as the measured signals depend on the shape of the pump pulse, which is very different for optical and THz excitation [Figs. 2(b,c)]. To solve this issue, we reshape the signals such that they refer to fictitious pump pulses with the same intensity profile. As the relevant signals [Figs. 2(b,c)] are quadratic in $E_\mathrm{p}$, they can be written as a convolution

$$\Delta S(t) = \left(H * E_\mathrm{p}^2\right)(t) = \int \mathrm{d}\tau\, H(t - \tau) E_\mathrm{p}^2(\tau) \tag{2}$$

of the pump intensity profile $E_\mathrm{p}^2(t)$ and the intrinsic response $H(t)$ of the system. Equation (2) implies that the two interactions with the pump field ($E_\mathrm{p}(\tau_1)$ and $E_\mathrm{p}(\tau_2)$) occur with negligible delay ($\tau_2 - \tau_1 = 0$). This assumption is reasonable because the memory time is given by the dephasing of electronic states at optical frequencies (~20 fs in metals [40]) and the transport relaxation time (~10 fs in our sample, see Fig. S4) for THz excitation.

Deconvolution of Eq. (2) is possible because the $E_\mathrm{p}^2(t)$ are precisely known. The procedure is detailed in the Supplemental Material and Fig. S5. It delivers a deconvoluted signal $\Delta S_\mathrm{dec}(t) = (E_\mathrm{fic}^2 * H)(t)$ that would be obtained for a fictitious pump pulse whose intensity profile $E_\mathrm{fic}^2$ is shown by the grey area in Fig. 3. By using this approach, we can push the time resolution of all signal traces down to the same value of 130 fs, even though the driving THz pulse is about 0.5 ps wide [Fig. 2(b)]. Because the signals $\Delta S_\mathrm{dec}$ are linear in $E_\mathrm{p}^2$, we scale the THz signals to match the optical traces at a delay of $t \approx 2$ ps, where electron and phonon baths should be almost fully thermalized [4].

Figure 3 shows that optical- and THz-pump-induced signals exhibit very similar dynamics. This remarkable finding is consistent with the raw signals of Fig. 2, whose shapes agree quite well, apart from differences during overlap of pump and probe pulses around $t = 0$. The additional oscillations seen on top of $\Delta S_{\text{dec}}$ (Fig. 3) are a consequence of the deconvolution procedure, which amplifies high-frequency noise components.

To summarize, optical and THz pump pulses are found to induce almost identical dynamics of the MOKE signal (Eq. (1)). Therefore, the dynamics of the magnetization is almost identical, too. This observation is surprising because (i) the number of initially photoexcited non-thermal electrons differ by three orders of magnitude (see above), and (ii) the distribution $\Delta n(\epsilon, t)$ of those carriers is strongly different (see Fig. 1). Thus, the data of Fig. 3 indicates that thermalization of electrons and their exact distribution after excitation do not play a dominant role in UDM on the scale of our time resolution of 130 fs.

**Model.** To understand this observation, we model UDM by spin-flip scattering of electrons [Fig. 4(a)]. Following Refs. [5,41], we assume that the dominant scattering events arise from impurities and phonons and are, therefore, approximately elastic. For an isotropic electron distribution, the occupation number $n^\sigma(\epsilon, t)$ of a spin-up ($\sigma = \uparrow$) or spin-down ($\sigma = \downarrow$) Bloch state only depends on its energy $\epsilon$ and on time $t$.

As indicated by Fig. 4(a), the rate of change of the electron occupation $n^\uparrow(\epsilon, t)$ of a spin-up Bloch state with energy $\epsilon$ due to elastic spin-flip scattering is proportional to $n^\uparrow D^\uparrow$ and the number $(1 - n^\downarrow)D^\downarrow$ of unoccupied spin-down states at the same energy $\epsilon$, minus an analogous term for the reverse process. Here, $D^\sigma(\epsilon)$ is the spin-dependent electronic density of states, and each transition is weighted with its squared matrix element $P(\epsilon)$. The rate of change in the magnetization $M \propto \int d\epsilon \left( D^\uparrow n^\uparrow - D^\downarrow n^\downarrow \right)$ is, therefore, given by

$$\dot{M} \propto -2 \int d\epsilon \left( \Delta n^\uparrow - \Delta n^\downarrow \right) D^\uparrow D^\downarrow P_{\text{sf}}, \tag{3}$$

where $\Delta n^\sigma(\epsilon, t)$ denotes the pump-induced change in the occupation number $n^\sigma$. In the linear fluence regime of our experiment (see Fig. S3), the weighting factor $\left( D^\uparrow D^\downarrow P_{\text{sf}} \right)(\epsilon, t)$ can be evaluated for the unexcited sample and is, thus, time-independent.

To explore Eq. (3), we consider two instructive scenarios right after excitation by the pump (photon energy $\hbar \omega_{\text{p}}$). First, we assume that $P_{\text{sf}}$, $D^\uparrow$ and $D^\downarrow$ are all independent of energy [Fig. 4(a)]. In the spin-up subset, the pump generates holes below and electrons above the Fermi level $\mu_0$. Because $P_{\text{sf}} D^\uparrow D^\downarrow$ is constant with respect to $\epsilon$, a spin-flip transition of an excited spin-up electron into a spin-down state is as likely as that of an excited spin-up hole. An analogous argumentation for the spin-down channel shows that no net magnetization change occurs. Therefore, an energy-dependent product $D^\uparrow D^\downarrow P_{\text{sf}}$ is essential to obtain a magnetization change.

Consequently, in the second scenario [Fig. 4(b)], we allow the spin-down density of states to depend on energy linearly, that is, $D^\downarrow(\epsilon) = D^\downarrow(\mu_0) + \gamma^\downarrow \cdot (\epsilon - \mu_0)$. From Fig. 4(b), we see that the excited spin-up electrons can scatter into a larger number of spin-down states than the excited spin-up holes. Therefore, the monotonic increase of $D^\downarrow(\epsilon)$ with energy $\epsilon$ enables magnetization quenching.

Note that the number of final spin-down states for spin-up electrons minus the analogous number for spin-up holes scales with $D^\downarrow(\epsilon + \hbar \omega_{\text{p}}) - D^\downarrow(\epsilon) = \gamma^\downarrow \cdot \hbar \omega_{\text{p}}$. On the other hand, the number of pump-excited spin-up electrons is proportional to $W_{\text{p}}^\uparrow / \hbar \omega_{\text{p}}$, where $W_{\text{p}}^\uparrow$ is the pump-pulse energy that is absorbed by spin-up electrons. Therefore, the impact of the pump photon energy $\hbar \omega_{\text{p}}$ cancels, and the demagnetization rate is only determined by $W_{\text{p}}^\uparrow$. This reasoning is illustrated in Fig. 4(b) for the two pump energies $\hbar \omega_{\text{p}}$ and $\hbar \omega_{\text{p}}/2$.

The preceding consideration can be extended to any pump-induced variation $\Delta n^\sigma$ of the electron distribution (see Supplemental Material and Ref. [7]). Assuming that no magnetization quenching has occurred yet ($\Delta M = 0$), we find that the quenching rate $\dot{M}$ is proportional to a weighted average of the energy the pump pulse has deposited in the spin-up and spin-down electrons, independent of the precise shape of the electron distribution (Supplemental Eqs. (31) and (33)).

More generally, one can assign generalized temperatures $T_0 + \Delta \tilde{T}^\sigma(t)$ to the instantaneous electron distributions $n_0 + \Delta n^\sigma(\epsilon, t)$. They reduce to the familiar temperatures when the two distributions $n_0 + \Delta n^\sigma$ are Fermi-Dirac functions (see Supplemental Eq. (21)). Assuming that the generalized temperatures of spin-up and spin-down electrons equilibrate rapidly, the electrons can be described by one common excess temperature $\Delta \tilde{T}_e = \Delta \tilde{T}^\uparrow = \Delta \tilde{T}^\downarrow$, which is directly proportional to the instantaneous excess energy of all electrons (see Supplemental Eq. (33)). In this case and for a $\delta(t)$-like pump pulse, the evolution of the magnetization can be shown (Supplemental Eq. (32)) to proceed according to

$$\Delta M(t) \propto \int_0^\infty d\tau \, e^{-\Gamma_{es}\tau} \Delta \tilde{T}_e(t - \tau), \tag{4}$$

where $\Gamma_{es}$ is the inverse time constant of electron-spin equilibration. If, for example, $\Delta \tilde{T}_e$ changes step-like, $\Delta M$ will adapt to the new temperature with time constant $\Gamma_{es}^{-1}$.

The dynamics of $\Delta \tilde{T}_e$ can be described by the familiar two-temperature model [42], even though $T_0 + \Delta \tilde{T}_e$ is a generalized temperature (see Supplementary Eq. (37)). We conclude that the dynamics of $\Delta \tilde{T}_e$ and, thus, $\Delta M$ are determined by the electronic excess energy, independent of the way it was deposited.

**Magnetization dynamics.** To extract the pure magnetization dynamics $\Delta \boldsymbol{M}(t) = \Delta M(t)\boldsymbol{u}_y$ from the transient MOKE rotation and ellipticity signals of Fig. 3, we use the procedure of Ref. [37] and the Supplemental Material. The resulting $\Delta M(t)/M_0$ curves are displayed in Fig. 5. As expected from the MOKE signals (Fig. 3), $\Delta M(t)$ proceeds almost identically for optical and THz excitation. The slight differences in the observed dynamics may appear due to uncertainties in the $\Delta M(t)$ extraction and deconvolution.

To compare the experimentally determined $\Delta M(t)$ with our model result (Eq. (4)), we assume that the dynamics of the electronic excess energy proceeds according to $\Delta \tilde{T}_e(t) \propto \Theta(t)[(1 - R)e^{-\Gamma_{ep}t} + R]$. Here, $\Theta(t)$ is the Heaviside step function, $\Gamma_{ep}^{-1}$ is the time constant of electron-phonon equilibration, and $R$ is the ratio of electronic and total heat capacity of the sample. With this input, Eq. (4) yields

$$\Delta M(t) \propto \Theta(t) \left[ \frac{\Gamma_{es} - R\Gamma_{ep}}{\Gamma_{es} - \Gamma_{ep}} e^{-\Gamma_{es}t} - \frac{(1-R)\Gamma_{es}}{\Gamma_{es} - \Gamma_{ep}} e^{-\Gamma_{ep}t} - R \right]. \tag{5}$$

To account for our non-$\delta(t)$-like pump pulse, Eq. (5) is convoluted with the pump-intensity profile $E_{fic}^2(t)$ (Fig. 3). We fit the resulting function to the measured $\Delta M(t)$ jointly for both THz and optical pump, where $\Gamma_{es}$, $\Gamma_{ep}$, $R$, a global scaling factor and a rigid time shift are free parameters. For the best fit (black solid curve in Fig. 5), we obtain $\Gamma_{es}^{-1} = 27 \pm 9$ fs, $\Gamma_{ep}^{-1} = 670 \pm 50$ fs and $R = 0.46 \pm 0.01$.

The value of $\Gamma_{es}^{-1}$ is approximately compatible with the electron-spin equilibration time found previously [4]. We note, however, that the value of $\Gamma_{es}^{-1}$ comes with significant uncertainty because it heavily relies on the procedure we used to extract $\Delta M(t)$ and because the duration of our fictitious pump pulse is significantly longer. The extracted value of $\Gamma_{ep}^{-1}$ is consistent with measurements of transient anisotropic reflectance $\Delta b(t)$ (see Eq. (1) and Fig. S6), whose decay is a monitor of electron-phonon equilibration [20], and with results from time-resolved photoelectron spectroscopy [43].

The value of $R$ obtained from the fit is much larger than the ratio of electronic and total heat capacity, which is about 0.06 for Fe [44]. This discrepancy is quite common for MOKE probing of UDM [4,5] and arises from the relatively large signal amplitude that remains after electrons and phonons have equilibrated ($t > 2$ ps in Figs. 3, 5). We attribute this effect to the pump-induced variation $\Delta a_y$ of the magneto-optic constants (Eq. (1)). The extraction procedure [37] of $\Delta M = \Delta M_y$ is designed to remove contributions of the increased electron temperature to $\Delta a_y$, which prevail directly after sample excitation. However, it likely does not cancel contributions due to the increased phonon temperature, which may become significant once electrons and phonons have thermalized.

**Discussion.** We observe almost identical magnetization dynamics of an Fe thin film following excitation by an optical and by a THz pump pulse (see Figs. 3, 5). This behavior is remarkable because the two pump pulses induce a very different transient state of the material. First, the THz pump pulse (photon energy of 4 meV) generates three orders of magnitude more electron-hole pairs than the optical pump (photon energy of 3.1 eV). Second, while excitation by the THz pulse merely results in a Fermi-Dirac electron distribution at elevated temperature, the optical pump induces a highly nonthermal electron distribution [Fig. 1(b)]. We are able to explain the identical dynamics following optical versus THz excitation by phase-space arguments for quasi-elastic spin-flip scattering [Fig. 4(b)].

Note that identical dynamics following optical versus THz excitation would also be observed if the optically induced nonthermal electron distribution relaxed (thermalized) to a Fermi-Dirac-type distribution much faster than the time resolution (~130 fs) of our experiment. This scenario appears, however, rather unlikely in view of known electron thermalization times $\Gamma_{ee}^{-1}$ of d-type metals such as Ni (~100 fs) [45,46], Fe (~100 fs) [47], Ru (~140 fs) [19], Pt (~100 fs) [20], Au (~1 ps) [48] or other metals like Gd (~100 fs) [18].

If electron thermalization was relevant for UDM, we would have to convolute the generalized temperature in Eq. (4) with a function that delays the rise of $\Delta\tilde{T}_e$ by $\Gamma_{ee}^{-1} = 100$ fs. To mimic this behavior, we first fit the THz-excitation data (for which electron thermalization is not relevant) in Fig. 5 with Eq. (5) and subsequently convolute the best fit by $H_{ee}(t) = \Gamma_{ee}\Theta(t)e^{-\Gamma_{ee}t}$. The resulting curve (black dotted line) drops significantly more slowly than the measured one and, thus, cannot describe the optical-excitation data. We conclude that electron thermalization is not relevant for UDM of Fe, consistent with our theoretical model (Eq. (4)).

Interestingly, the difference between rotation and ellipticity signals shows that pump-induced changes $\Delta\boldsymbol{a}(t)$ in the magneto-optic coefficients contribute significantly to the measured MOKE signal, also after deconvolution (Fig. 3). This observation and the identical observed dynamics of the MOKE signals following optical and THz excitation (Fig. 3) suggest that $\Delta\boldsymbol{a}(t)$ also scales with the electronic excess energy, independent of the shape of the electron and, possibly, phonon distribution. An analogous statement can be made for the dynamics of the non-magnetic signal contribution $\Delta b(t)$ (see Eq. (1) and Fig. S7).

We emphasize that optical and THz excitation of metals do not always induce identical spin dynamics. An important example is the ultrafast spin Seebeck effect in stacks of yttrium iron garnet (YIG) and Pt thin films. Excitation of the Pt layer leads to a spin current from YIG to Pt, whose magnitude is the largest when the excited electron distribution in Pt has relaxed to a Fermi-Dirac distribution [20].

**Conclusions.** We have shown that both optical and THz excitation of a ferromagnetic Fe thin film result in identical magnetization dynamics, even though the two pulses induce very different changes in the electron distribution directly after absorption. Phase-space considerations of quasi-elastic spin-flip scattering can consistently explain our observations. They show that the UDM dynamics are determined by the electronic excess energy, independent of the way it was deposited.

Our results are also interesting from an applied viewpoint. First, the speed of UDM is not limited by carrier multiplication [20]. Second, because ultrafast spin transport is driven by the same force as UDM [6,7], it should exhibit the same dynamics, independent of the photon energy of the pump pulse. Such behavior was indeed reported recently [10,11].

**Supplemental material**

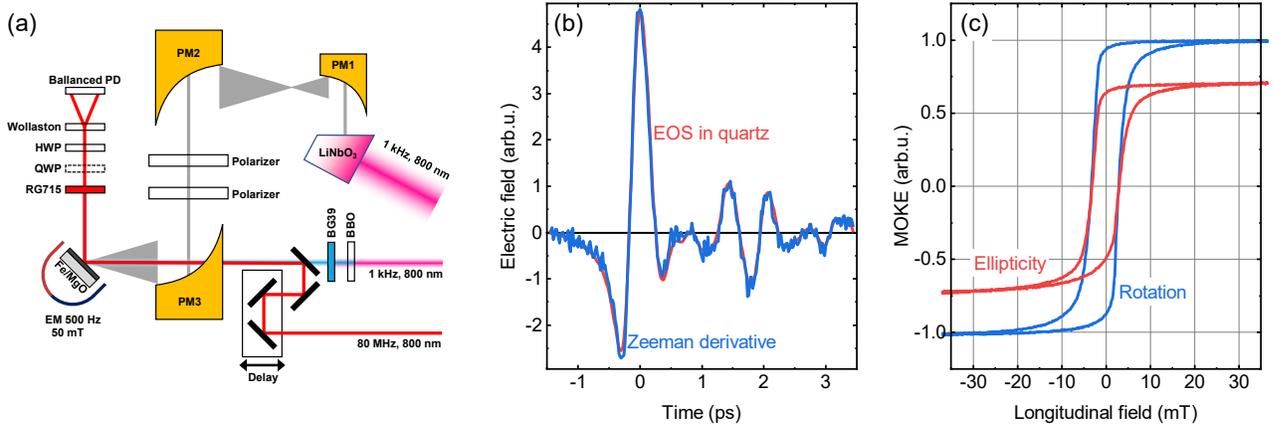

Figure S1. (a) Experimental setup for measuring UDM with optical pump pulses (mean photon energy 3.1 eV) and terahertz pump pulses (4.1 meV). (b) Electric field of a terahertz pulse measured by electro-optic sampling (EOS) in quartz and corrected for the quartz response function (red) and derived from the Zeeman contribution to the ultrafast MOKE signal (blue). (c) Hysteresis loops of MOKE rotation and ellipticity with the magnetization along the $y$ axis [see Fig. 1(a)].

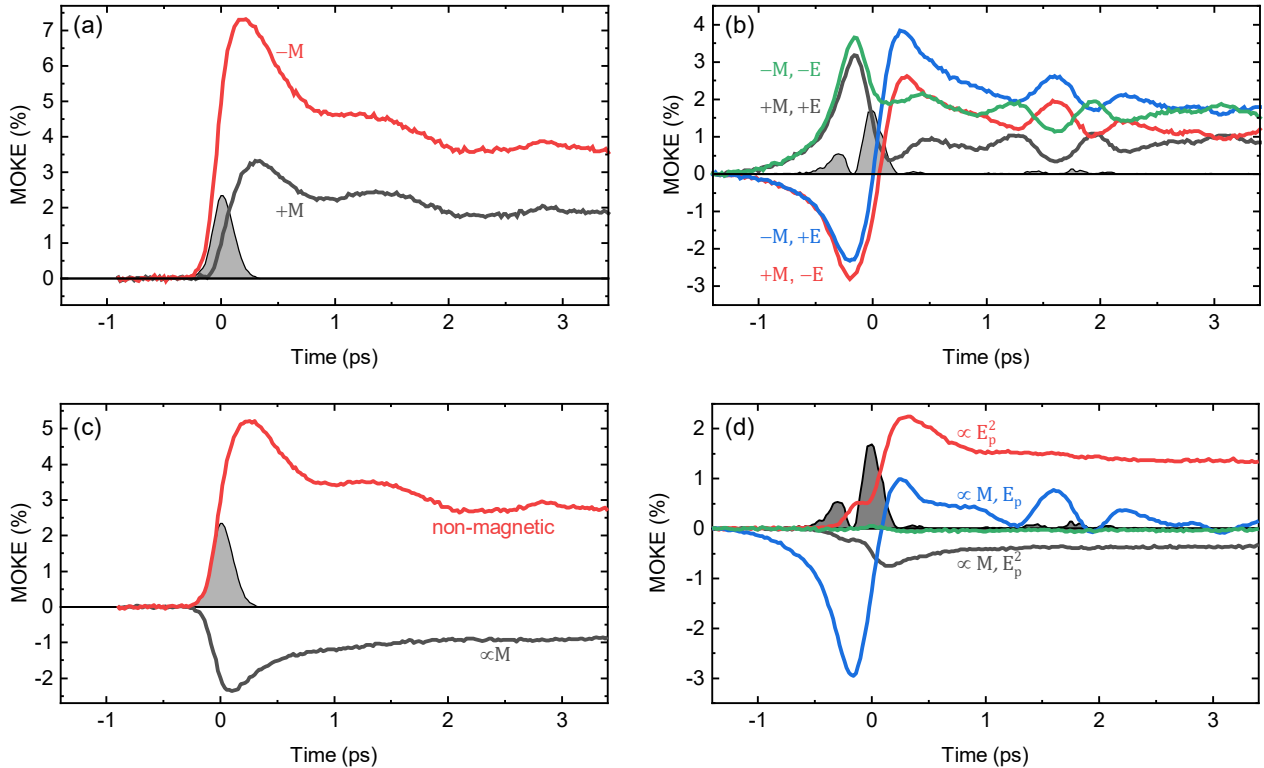

Figure S2. Raw signals measured for all possible combinations of magnetization and terahertz field directions for (a) optical and (b) terahertz pump. By suitable linear combinations of the traces, we separate different contributions in the dynamics after (c) optical and (d) terahertz excitation.

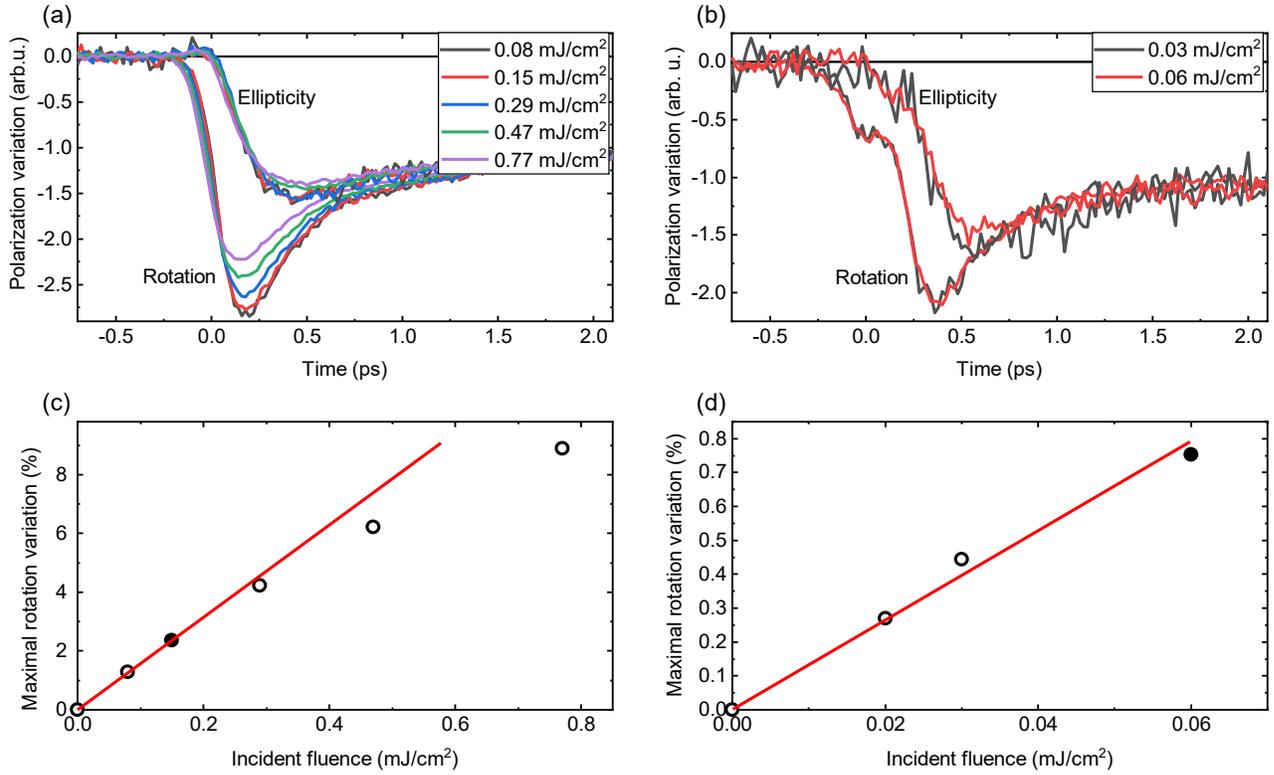

Figure S3. (a,b) Normalized rotation and ellipticity signals odd in $M_0$ for several incident fluences of optical pump [panel (a)] and THz pump [panel (b)]. (c,d) Maximum transient MOKE rotation as a function of incident pump fluence for optical [panel (c)] and THz excitation [panel (d)]. Red lines are guide to the eye, and filled symbols indicate fluences chosen for the analysis.

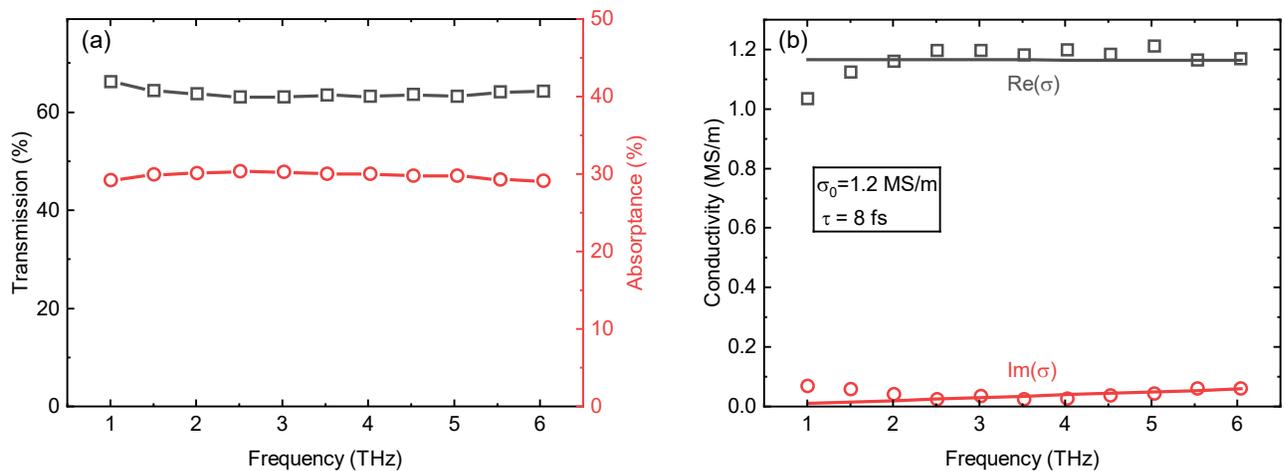

Figure S4. (a) Measured THz transmission $T(\omega)$ of the sample (black) for normal incidence and extracted absorptance $A_{\text{THz}}(\omega)$. (b) THz conductivity $\sigma(\omega)$ fitted with the Drude model, yielding electron scattering time of ~8 fs.

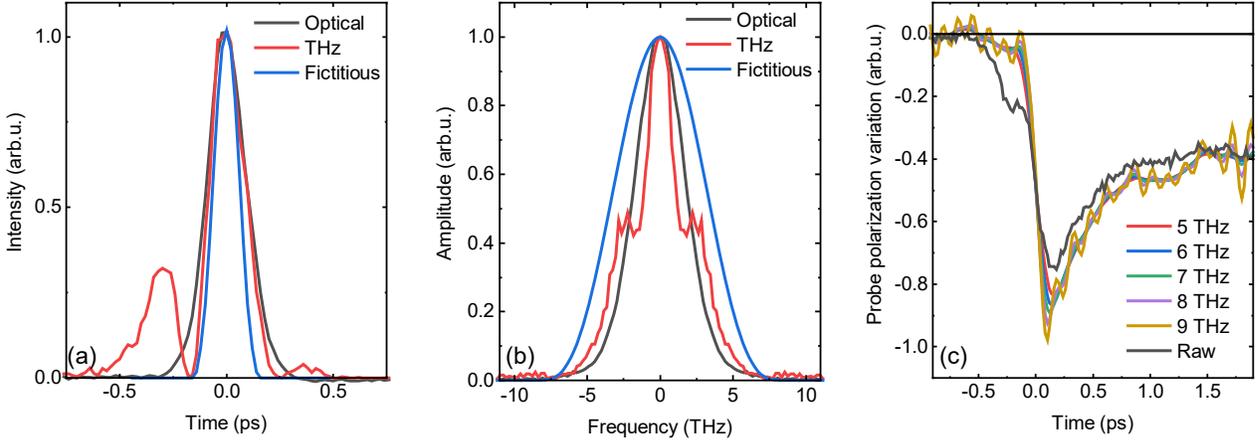

Figure S5. Normalized pump pulses in (a) time and (b) frequency domain for optical (black), THz (red) and fictitious pump pulse (blue). The cut-off frequency of $E_{\text{fic}}^2(t)$ is set to 8 THz. (c) Deconvoluted odd-in-$M_0$ MOKE rotation signal for THz pump and different cut-off frequencies together with raw signal (black).

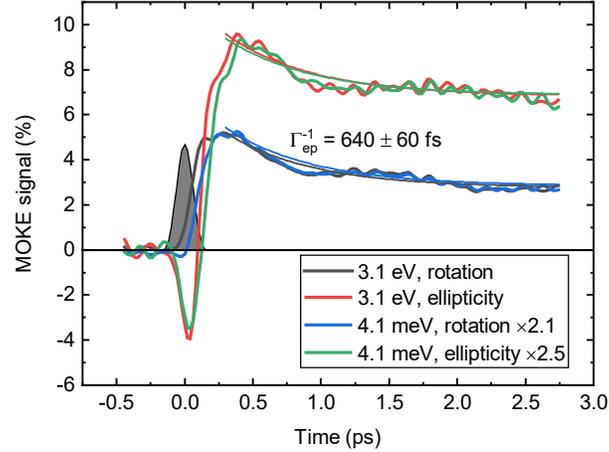

Figure S6. MOKE signals even in $M_0$ and $E_p$ after deconvolution for optical and THz pump. They correspond to $\Delta b(t)$ of Eq. (1). A joint exponential fit after 0.4 ps provides the value of the electron-phonon thermalization time $\Gamma_{\text{ep}}^{-1}$, which is very similar to the one obtained from magnetic signals (Fig. 5).

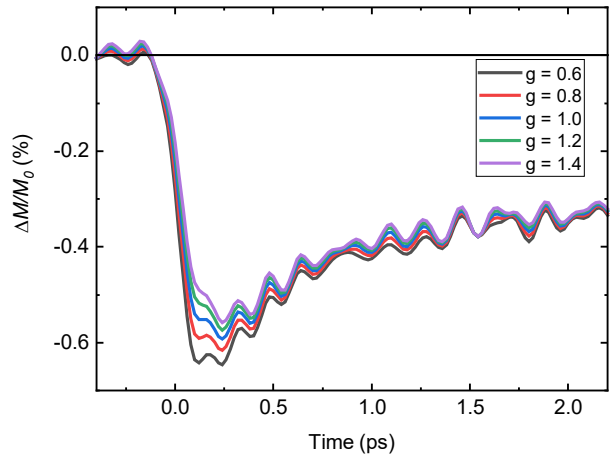

Figure S7. Extraction of $\Delta M(t)$ from MOKE rotation and ellipticity signals for different parameter $g$ values.

**Setup details.** For generation of the pump pulses, we use an amplified Ti:sapphire system. Optical pump pulses (spot diameter 200 µm, photon energy 3.1 eV, repetition rate 1 kHz, p-polarized) are generated by frequency doubling of the laser output in the BBO crystal. The THz pulses (900 µm, 4.1 meV, 1 kHz, p-polarized) are generated by tilted-wavefront-type optical rectification in a LiNbO$_3$ crystal and focused on the sample using three parabolic mirrors (see Fig. S1a). For the probe pulses, we use the output of the amplified system seed oscillator (100 µm, 1.6 eV, 80 MHz, polarized at 3° out of the s axis).

The paths of probe and optical pump are combined by means of a dichroic mirror. The probe is reflected from the magnetized sample [Fig. 1(a)] and directed into the balanced detection. An optional quarter waveplate serves to switch between measurements of probe rotation and ellipticity. The polarity of the external magnetic field (magnitude 40 mT) is alternated at a rate of 500 Hz. We use two THz polarizers to alternate the polarity of the THz pulse. A set of filters serves to block the fundamental after the BBO crystal and to block the optical pump after reflection from the sample.

**Sample details.** The Fe thin-film was grown by electron-beam evaporation at a rate of $0.2 \text{ Å s}^{-1}$ and a base pressure in the $10^{-9}$ mbar range. The film has in-plane magnetic anisotropy. Hysteresis loops of $M_y$ as measured by the longitudinal MOKE are shown in Fig. S1(c).

**THz transmission spectroscopy.** The THz conductance of the Fe thin film was determined by THz transmission spectroscopy [49,50], in which we measure THz fields $E_{\text{Fe}||\text{sub}}$ and $E_{\text{sub}}$ through the sample Fe||sub (Fe film on substrate) and the substrate (sub) without Fe film, respectively. In the frequency domain, the ratio $T(\omega)$ of the two fields obeys

$$T(\omega) = \frac{E_{\text{Fe}||\text{sub}}(\omega)}{E_{\text{sub}}(\omega)} = \frac{n_a(\omega) + n_{\text{sub}}(\omega)}{n_a(\omega) + n_{\text{sub}}(\omega) + Z_0 G(\omega)}, \quad (6)$$

where $n_a(\omega) \approx 1$ and $n_{\text{sub}}(\omega) \approx 2.1$ are the known refractive indices of air and the glass substrate [51], respectively [see Fig. 2(a)], and $Z_0 \approx 377\ \Omega$ is the free-space impedance. The conductance $G(\omega) = \sigma(\omega)d$ of the nominally homogeneous Fe film is given by the product of its conductivity $\sigma(\omega)$ and thickness $d = 4$ nm. We, thus, obtain

$$G(\omega) = \frac{n_a(\omega) + n_{\text{sub}}(\omega)}{Z_0}\left[\frac{1}{T(\omega)} - 1\right], \quad (7)$$

from which the conductivity $\sigma(\omega) = G(\omega)/d$ is obtained. Results of the THz transmission measurements are shown in Fig. S4. By fitting the conductivity with a Drude model [see Fig. S4(b)], we obtain an electron velocity relaxation time of approximately 8 fs.

**Processing of raw signals.** The measured pump-induced signals are probe rotation/ellipticity changes [see Fig.S1(a)] with respect to the equilibrium value, when no pump is applied $\Delta S(t) = S(t) - S_0$. The signals are recorded for two opposite magnetizations $\pm \mathbf{M}_0$ of the Fe film and two opposite pump fields $\pm \mathbf{E}_p(t)$ in the case of THz pumping [Fig. S2(a,c)]. To separate various contributions by their parity with respect to $\mathbf{M}_0$ and $\mathbf{E}_p$, we take corresponding sums (even parity) and differences (odd parity) of these signals. For example, to separate the contribution linear in $\mathbf{M}_0$ and quadratic in $\mathbf{E}_p$ (which corresponds to UDM), we use

$$4\Delta S^{\mathbf{M}_0, E_p^2}(t) = \Delta S(t, +\mathbf{M}_0, +\mathbf{E}_p) - \Delta S(t, -\mathbf{M}_0, +\mathbf{E}_p) + \Delta S(t, +\mathbf{M}_0, -\mathbf{E}_p) - \Delta S(t, -\mathbf{M}_0, -\mathbf{E}_p) \quad (8)$$

and similarly for all other contributions [Fig. S2(b,d)].

The signals corresponding to UDM can be normalized by the height of the MOKE hysteresis loop given by $S_0(\mathbf{M}_0) - S_0(-\mathbf{M}_0) = 2a_{0y}M_0$. Based on Eq. (1), such normalization yields relative variations of magnetization and magneto-optical constant:

$$\frac{2\Delta S^{M_0,E_p^2}(t)}{S_0(+M_0) - S_0(-M_0)} = \frac{a_0}{a_{0y}} \cdot \frac{\Delta M(t)}{M_0} + \frac{\Delta a_y(t)}{a_{0y}} = \frac{\Delta M_y(t)}{M_0} + \frac{\Delta a_y(t)}{a_{0y}}. \tag{9}$$

Similar to the odd-in-$M_0$ responses, we perform the analysis for the non-magnetic responses (even in $M_0$ and $E_p$) and directly compare them for optical and THz pump.

**Absorbed fluences.** To determine the THz absorptance $A_{\text{THz}}(\omega)$ of the Fe film, we take advantage of the THz transmission measurements. The absorbed power per volume is given by $dP_{\text{abs}}/dV = |E_{\text{Fe}}(\omega)|^2 \operatorname{Re}\sigma(\omega)/2$. The approximately homogeneous THz pump field $E_{\text{Fe}}(\omega)$ within and directly behind the Fe film is related to the incident field $E_{\text{inc}}(\omega)$ and the measured $T(\omega)$ (Eq. (6)) through $E_{\text{Fe}}(\omega) = 2n_a(\omega)T(\omega)E_{\text{inc}}(\omega)/[n_a(\omega) + n_{\text{sub}}(\omega)]$, whereas $\sigma(\omega) = G(\omega)/d$ is obtained from Eq. (7). Integration of $dP_{\text{abs}}/dV$ from $z = 0$ to $d$ yields the absorbed power per sample area,

$$\frac{dP_{\text{abs}}}{dA} = \frac{|E_{\text{inc}}(\omega)|^2}{2Z_0} \left| \frac{2n_a(\omega)T(\omega)}{n_a(\omega) + n_{\text{sub}}(\omega)} \right|^2 \operatorname{Re}\left\{ \left[\frac{1}{T(\omega)} - 1\right][n_a(\omega) + n_{\text{sub}}(\omega)] \right\}. \tag{10}$$

By assuming a real-valued $n_a(\omega)$, the power flux $dP_{\text{inc}}/dA$ of the incident wave can be written as $n_a(\omega)|E_{\text{inc}}(\omega)|^2/2Z_0$, and with Eq. (10), its absorbed fraction becomes

$$A_{\text{THz}}(\omega) = \frac{dP_{\text{abs}}/dA}{dP_{\text{inc}}/dA} = n_a(\omega) \left| \frac{2T(\omega)}{n_a(\omega) + n_{\text{sub}}(\omega)} \right|^2 \operatorname{Re}\left\{ \left[\frac{1}{T(\omega)} - 1\right][n_a(\omega) + n_{\text{sub}}(\omega)] \right\}. \tag{11}$$

We obtain $A_{\text{THz}} \approx 0.3$ for normal incidence [Fig. S4(a)].

Because the pump-induced MOKE signal $\Delta S^{M_0,E_p^2}$ depends linearly on the pump fluence (see Fig. S3), the ratio of the pump-probe signals long after excitation should be equal to the ratio of absorbed fluences $F_{\text{opt}}A_{\text{opt}}/F_{\text{THz}}A_{\text{THz}}$, where $F_{\text{opt}}, F_{\text{THz}}$ is the incident fluence, and $A_{\text{opt}}, A_{\text{THz}}$ is the absorptance of the Fe film. From the incident powers and approximate pump-spot sizes, we calculate incident fluences of $F_{\text{opt}} = 0.15 \text{ mJ cm}^{-2}$ and $F_{\text{THz}} = 0.06 \text{ mJ cm}^{-2}$. We estimate the absorption coefficients for the case of normal incidence and obtain $A_{\text{Opt}} \approx 0.4$ [9] and $A_{\text{THz}} \approx 0.3$ (see above). The result is $F_{\text{opt}}A_{\text{opt}}/F_{\text{THz}}A_{\text{THz}} \approx 3.6$, which is in reasonable agreement with the experimental value of 2.7 (see Fig. 3).

**Deconvolution procedure.** For the optical pump, we measure $E_p^2(t)$ through the optical Kerr effect in a 300 μm thick diamond window. For the THz pump, we calculate $E_p^2(t) \propto B_p^2(t)$ from the transient THz magnetic field $B_p(t)$, which is obtained from the Zeeman-torque contribution. The response function $H(t)$ is obtained by Fourier transformation of Eq. (2) and solving for $H(\omega)F(\omega)$, where $F(\omega)$ is the Fourier transformation of the fictitious pump-pulse intensity $E_{\text{fic}}^2(t)$ and set to zero above a cut-off frequency of 8 THz [Fig. S5(a)]. Therefore, $F(\omega)$ acts like a low-pass filter [Fig. S5(b)]. Results of the deconvolution procedure are shown in [Fig. S5(c)] for various cut-off frequencies of $F(\omega)$.

**Extracting magnetization dynamics.** For $\Delta M_x = \Delta M_z = 0$, the relative pump-induced change in the MOKE signal can according to Eq. (9) be rewritten as

$$\delta S(t) = \delta M(t) + \delta a(t), \tag{12}$$

where $\delta S(t)$ equals the left-hand side of Eq. (9), $\delta M(t) = \Delta M_y/M_0$ and $\delta a(t) = \Delta a_y/a_{y0}$. To separate the components $\delta M(t)$ and $\delta a(t)$ of our signal, we adopt the method used in Ref. [37]. As seen in Fig. 2, the contribution $\delta a(t)$ leads to different dynamics of rotation and ellipticity signals because in general, $\delta a^{\text{rot}}(t) \neq \delta a^{\text{ell}}(t)$. We assume that $\delta a^{\text{rot}}(t) = g\delta a^{\text{ell}}(t)$, where $g$ is a proportionality constant. We have, thus, three equations for three unknowns, which yield

$$\delta M(t) = \frac{\delta S^{\text{rot}}(t) - g\delta S^{\text{ell}}(t)}{1 - g}. \tag{13}$$

As in Ref. [37], we choose the value of the unknown constant $g$ such that the resulting signal $\delta M(t)$ does not exhibit discontinuities in the vicinity of $t = 0$, that is, of the arrival time of the pump. Interestingly, the optimal value of $g$ is approximately -1 for both optical and terahertz pump, thereby resulting in simple summation of the two transients (see Eq. (13)). The resulting magnetization transients are shown in Fig. 5. The impact of various values of $g$ is shown in Fig. S7.

**Theory: rate of magnetization change.** We assume that the state of the electronic system is fully characterized by the occupation numbers $n_k^\sigma$ of a Bloch state $(k, \sigma)$ where $\sigma = \uparrow, \downarrow$ refers to the electron spin, and $k$ summarizes the band index and wavevector. We also assume that the $k$-dependence is fully given by the energy $\epsilon_k^\sigma$ of the Bloch state, that is,

$$n_k^\sigma(t) = n^\sigma(\epsilon_k^\sigma, t). \tag{14}$$

In the framework of the Stoner model, the Bloch energy depends on the pump-induced change $\Delta m$ in the magnetic moment according to

$$\epsilon_k^\sigma(t) = \epsilon_{k0}^\sigma + I^\sigma \Delta m(t). \tag{15}$$

Here, $\epsilon_{k0}^\sigma$ is the Bloch energy before arrival of the pump pulse, and $I^\sigma = I^{\uparrow,\downarrow} = \pm I/2$ quantifies the strength of the effective electron-electron Coulomb interaction.

Note that we are only interested in effects up to first order in the pump-induced changes

$$\Delta n^\sigma = n^\sigma - n_0 \tag{16}$$

in the distribution functions. Before arrival of the pump pulse, the $n^\sigma(\epsilon, t)$ are given by one and the same Fermi-Dirac function $n_0(\epsilon)$ at temperature $T_0$. The reason is that in our experiments, all demagnetization signals are found to scale linearly with the deposited pump-pulse energy.

The magnetic moment of the sample in Bohr magnetons is given by

$$m = \int d\epsilon \left( D^\uparrow n^\uparrow - D^\downarrow n^\downarrow \right). \tag{17}$$

As derived in the main text, the rate of change in $m$ is given by Eq. (3), which is rewritten as

$$\Delta \dot{m} = -2 \int d\epsilon \left( \Delta n^\uparrow - \Delta n^\downarrow \right) g_{\text{sf}}. \tag{18}$$

Here, $g_{\text{sf}}(\epsilon) = \left( P_{\text{sf}} D^\uparrow D^\downarrow \right)(\epsilon)$ scales the probability that a spin-up electron with energy $\epsilon$ makes a transition into spin-down states of the same energy. Because the density $D^\uparrow$ and $D^\downarrow$ of, respectively, spin-up and spin-down electronic states and the squared matrix $P_{\text{sf}}$ element can all depend on the Bloch-electron energy $\epsilon$, the product $g_{\text{sf}}$ does also.

To analyze Eq. (18), we assume that the weight factor $W = g_{\text{sf}}$ can be approximated by

$$W(\epsilon) = W(\mu_0) + W'(\mu_0)(\epsilon - \mu_0) \tag{19}$$

for all energies $\epsilon$ around the unperturbed chemical potential $\mu_0$ at which the changes $\Delta n^\uparrow$ and $\Delta n^\downarrow$ in the electron distribution are notably nonzero. The integral over $W \Delta n^\sigma$ ($\sigma = \uparrow, \downarrow$) then yields

$$\int d\epsilon \, W \Delta n^\sigma = W(\mu_0) \Delta P^\sigma + W'(\mu_0) \Delta A^\sigma \tag{20}$$

which is just a linear combination of the zeroth and first moment of the distribution changes $\Delta n^\sigma$. The moments are defined as

$$\Delta P^\sigma = \int d\epsilon \, \Delta n^\sigma \quad \text{and} \quad \Delta A^\sigma = \frac{\pi^2 k_B^2}{3} T_0 \Delta \tilde{T}^\sigma = \int d\epsilon \, (\epsilon - \mu_0) \Delta n^\sigma. \tag{21}$$

In the case when $n^\sigma = n_0 + \Delta n^\sigma$ is a Fermi-Dirac distribution with chemical potential $\mu^\sigma$ and temperature $T^\sigma$, $\Delta P^\sigma$ and $\Delta A^\sigma$ reduce to

$$\Delta P^\sigma = \mu^\sigma - \mu_0 \quad \text{and} \quad \Delta \tilde{T}^\sigma = T^\sigma - T_0 \tag{22}$$

to linear order in the $\Delta n^\sigma$. One can, thus, interpret $\Delta P^\sigma$ and $\Delta \tilde{T}^\sigma \propto \Delta A^\sigma$ as changes in, respectively, a generalized chemical potential and a generalized temperature. We emphasize, however, that the definition of the moments $\Delta P^\sigma$ and $\Delta A^\sigma$ (Eq. (21)) also applies to nonthermal electron distributions $n_0 + \Delta n^\sigma$. We will use both $\Delta A^\sigma$ and $\Delta \tilde{T}^\sigma$ for convenience.

By applying the moment expansion of Eq. (20) to Eq. (18), we obtain

$$\Delta \dot{m} = -2g_{sf}(\mu_0)\Delta P_s - 2g'_{sf}(\mu_0)(\Delta A^\uparrow - \Delta A^\downarrow). \tag{23}$$

The first term on the right-hand side describes magnetization relaxation driven by the spin voltage

$$\Delta P_s = \Delta P^\uparrow - \Delta P^\downarrow. \tag{24}$$

The term proportional to $\Delta A^\uparrow - \Delta A^\downarrow$ is a term analogous to the Seebeck effect, which contributes as long as the generalized temperatures of spin-up and spin-down electrons are different.

**Magnetization dynamics.** To eliminate $\Delta P_s$ in Eq. (23), we take advantage of particle conservation. The change in the sample magnetic moment (Eq. (17)) is given by $\Delta m = \Delta(N^\uparrow - N^\downarrow)$, where $N^\sigma = \int d\epsilon\, D^\sigma n^\sigma$ is the number of electrons with spin $\sigma = \uparrow, \downarrow$. To linear order, the pump-induced change is

$$\Delta N^\sigma = \int d\epsilon\, (\Delta D^\sigma n_0 + D_0^\sigma \Delta n^\sigma). \tag{25}$$

The pump-induced change $\Delta D^\sigma$ in the density of states arises from the magnetic-moment change through Eq. (15). We express $\Delta D^\sigma(\epsilon)$ to first order in the $\Delta n^\sigma$ by $\Delta D^\sigma(\epsilon) = D_0^{\sigma\prime}(\epsilon)I^\sigma \Delta m$. By expanding $D_0^\sigma(\epsilon) = D_0^\sigma(\mu_0) + D_0^{\sigma\prime}(\mu_0)(\epsilon - \mu_0)$ (see Eq. (19)), we obtain $\int d\epsilon\, \Delta D^\sigma n_0 = D_0^\sigma(\mu_0)I^\sigma \Delta m$ for the first term of Eq. (25). By applying the moment expansion (Eq. (20)) to the second term, we find

$$\Delta N^\sigma = D_0^\sigma(\mu_0)I^\sigma \Delta m + D_0^\sigma(\mu_0)\Delta P^\sigma + D_0^{\sigma\prime}(\mu_0)\Delta A^\sigma. \tag{26}$$

By taking particle conservation into account, $\Delta(N^\uparrow + N^\downarrow) = 0$, and making use of $\Delta m = \Delta(N^\uparrow - N^\downarrow)$ and $I^\sigma = \pm I/2$, Eq. (26) becomes

$$\pm(1 - ID_0^\sigma(\mu_0))\Delta m = 2D_0^\sigma(\mu_0)\Delta P^\sigma + 2D_0^{\sigma\prime}(\mu_0)\Delta A^\sigma \tag{27}$$

for $\sigma = \uparrow, \downarrow$. As we are interested in the relationship with the spin chemical potential $\Delta P_s = \Delta P^\uparrow - \Delta P^\downarrow$, we subtract the two Eqs. (27) and finally find

$$\frac{1}{\chi_s}\Delta m = \Delta P_s + \frac{D_0^{\uparrow\prime}}{D_0^\uparrow}(\mu_0)\Delta A^\uparrow - \frac{D_0^{\downarrow\prime}}{D_0^\downarrow}(\mu_0)\Delta A^\downarrow \tag{28}$$

where the Pauli spin susceptibility $\chi_s > 0$ is given by

$$\frac{1}{\chi_s} = \frac{1}{2D_0^\uparrow(\mu_0)} + \frac{1}{2D_0^\downarrow(\mu_0)} - I. \tag{29}$$

We have, thus, expressed $\Delta m$ by the moments $\Delta P_s$ and $\Delta A^\sigma$. Note that in standard thermodynamics with state variables $\mu_s = \mu^\uparrow - \mu^\downarrow$ and one temperature $T$, Eq. (28) corresponds to

$$\Delta m = \frac{\partial m}{\partial \mu_s}\Delta \mu_s + \frac{\partial m}{\partial T}\Delta T. \tag{30}$$

From Eq. (28), we can substitute $\Delta P_s$ into Eq. (23) and obtain the equation of motion for $\Delta m$,

$$(\partial_t + \Gamma_{es})\Delta m = \Delta F = 2g_{sf}\left(\frac{D_0^{\uparrow\prime}}{D_0^{\uparrow}} - \frac{g_{sf}'}{g_{sf}}\right)(\mu_0)\Delta A^{\uparrow} - 2g_{sf}\left(\frac{D_0^{\downarrow\prime}}{D_0^{\downarrow}} - \frac{g_{sf}'}{g_{sf}}\right)(\mu_0)\Delta A^{\downarrow}, \quad (31)$$

where $\Gamma_{es} = 2g_{sf}(\mu_0)/\chi_s$, and $\Delta F$ is the force that drives the magnetization change. It is proportional to a weighted average of the generalized excess electron temperatures.

The solution of Eq. (30) is given by the convolution

$$\Delta m(t) = (H_{es} * \Delta F)(t) = \int d\tau\, H_{es}(t-\tau)\Delta F(\tau) \quad (32)$$

of $\Delta F$ with a response function $H_{es}(t) = \Theta(t)e^{-\Gamma_{es}t}$, where $\Theta(t)$ is the Heaviside step function. The time constant $\Gamma_{es}^{-1}$ quantifies how quickly the magnetization adapts to a change in the squared generalized temperatures of spin-up and spin-down electrons.

In the framework of the Stoner model, the energy of the electrons is given by $E_e = \sum_\sigma \int d\epsilon\,(\epsilon - \mu_0)D^\sigma n^\sigma + Im^2/4$. It is modified by excitation by the laser-pulse and electron-phonon interactions through changes $\Delta n^\sigma$, the sample magnetic moment $m$ and, thus, the density $D^\sigma$ of states. By a consideration similar to Eq. (26), one can show that the latter two processes cancel to linear order in the $\Delta n^\sigma$, and one obtains

$$\Delta E_e = \sum_\sigma C_e^\sigma \Delta \tilde{T}^\sigma \quad (33)$$

for the electron excess energy. Here, $C_e^\sigma = (\pi^2 k_B^2/3)T_0 D_0^\sigma(\mu_0)$ and $\Delta \tilde{T}^\sigma \propto \Delta A^\sigma$ (Eq. (21)). Note that $C_e = C_e^\uparrow + C_e^\downarrow$ is the heat capacity of the electrons.

Equation (33) underscores the interpretation of $T_0 + \Delta \tilde{T}^\sigma$ as generalized temperature (Eq. (21)) and allows us to rephrase the interpretation of Eq. (31): Once spin-up and spin-down electrons have equilibrated ($\Delta \tilde{T}^\uparrow = \Delta \tilde{T}^\downarrow$), $\Delta F$ is directly proportional to the excess energy of all electrons, independent of the precise shape of the electron distributions.

**Energy transfer to phonons.** After excitation by the pump pulse, the excess electron energy $\Delta E_e$ decays due to electron-phonon (ep) interactions. To determine the dynamics of $\Delta E_e$, we neglect spin flips and use the relationship [42]

$$\Delta \dot{E}_e\big|_{ep} \propto \sum_\sigma \int d\delta\,(\alpha^2 F)^\sigma(\delta)\int d\epsilon\,\{[n^\sigma(\epsilon) - n^\sigma(\epsilon+\delta)]p(\delta) - [1 - n^\sigma(\epsilon)]n^\sigma(\epsilon+\delta)\}. \quad (34)$$

Here, $(\alpha^2 F)^\sigma(\delta)$ denotes the Eliashberg function that describes the coupling of phonons of energy $\delta$ with two electronic states of the same spin $\sigma$ and energy $\epsilon$ and $\epsilon + \delta$. The occupation of phonons with energy $\delta$ is given by $p(\delta)$. Note that the term under the $\epsilon$-integral becomes zero for all $\delta$ and $\epsilon$ if $n^\sigma$ is a Fermi-Dirac distribution and $p$ is a Bose-Einstein distribution with the same temperature.

By linearizing Eq. (34) with respect to $\Delta n^\sigma = n^\sigma - n_0$ and $\Delta p = p - p_0$, we obtain

$$\Delta \dot{E}_e\big|_{ep} \propto \sum_\sigma \int d\delta\,(\alpha^2 F)^\sigma(\delta)\delta\Delta p(\delta) - \sum_\sigma \int d\epsilon\,\Delta n^\sigma(\epsilon)\int d\delta\,(\alpha^2 F)^\sigma[1 - n_0(\epsilon-\delta) - n_0(\epsilon+\delta)]. \quad (35)$$

Because the weight factor of $\Delta n^\sigma(\epsilon)$ in Eq. (35) is sufficiently smooth, it is legitimate to apply the moment expansion of Eq. (20), resulting in

$$\Delta \dot{E}_e\big|_{ep} \propto \sum_\sigma \int d\delta\,(\alpha^2 F)^\sigma(\delta)\delta\Delta p(\delta) - \sum_\sigma \Delta A^\sigma \int d\delta\,(\alpha^2 F)^\sigma(\delta)[-2n_0'(\mu_0 - \delta)] \quad (36)$$

Note that the first integral approximately scales with the pump-induced phonon excess energy because $(\alpha^2 F)^\sigma(\delta)$ is approximately proportional to the phonon density of states [42]. Owing to Eqs. (21) and (33), the second integral approximately scales with the excess energy of the $\sigma$-electrons. The generalized chemical potential does not show up in Eq. (36) as the weight factor of $\Delta n^\sigma(\epsilon)$ in Eq. (35) is antisymmetric with respect to $\epsilon - \mu_0$.

When we finally assume that the phonon distribution $p_0 + \Delta p$ is thermal and obeys a Bose-Einstein distribution at temperature $T_0 + \Delta T_p$, Eq. (36) along with Eqs. (21) and (33) lead to the familiar result

$$C_e \, \partial_t \Delta \tilde{T}_e \big|_{ep} = -G_{ep} \cdot (\Delta \tilde{T}_e - \Delta T_p). \tag{37}$$

Here, $C_e = C_e^\uparrow + C_e^\downarrow$ is the total electronic heat capacity, and the coupling strength $G_{ep}$ is proportional to $\sum_\sigma \int d\delta \, (\alpha^2 F)^\sigma(\delta)[-2n_0'(\mu_0 - \delta)]$. In the last step to Eq. (37), we took advantage of the fact that $\Delta \dot{E}_e \big|_{ep} = 0$ when $\Delta \tilde{T}_e = \Delta T_p$. We emphasize that the electron-phonon energy-transfer rate is given by the difference between the generalized temperature of the $\sigma$-electrons and the thermal phonons, independent of the precise shape of the possibly nonthermal electron distribution.

**Distribution function after pump absorption.** The change $\Delta n(\epsilon)$ in the electron occupation number directly after absorption of a pump pulse with photon energy $\hbar \omega_p$ is proportional to the strength of transitions from states with energy $\epsilon \pm \hbar \omega_p$, that is,

$$\Delta n(\epsilon) = C(\epsilon)\{[n_0(\epsilon - \hbar \omega_p) - n_0(\epsilon)] + [n_0(\epsilon + \hbar \omega_p) - n_0(\epsilon)]\}. \tag{38}$$

Here, $n_0(\epsilon)$ is a Fermi-Dirac distribution with temperature $T = T_0$ and chemical potential (Fermi level) $\mu_0$. The prefactor $C(\epsilon)$ captures matrix elements, densities of states and the pump-pulse energy and is assumed to be constant over the relevant electron energies. The absorbed pump energy is given by

$$\Delta W_{abs} = \Delta E_e = \int d\epsilon \, (\epsilon - \mu_0) D_0(\epsilon) \Delta n(\epsilon) \approx D_0(\mu_0) \int d\epsilon \, (\epsilon - \mu_0) \Delta n(\epsilon), \tag{39}$$

where $D_0 = D_0^\uparrow + D_0^\downarrow$ is the electronic density of states before arrival of the pump pulse. In the last step of Eq. (39), we applied the moment expansion of Eq. (20). The density of states can be obtained from the known molar electronic specific heat $C_e/N = \gamma T_0$ of Fe, where $N$ is the number of Fe atoms and $\gamma \approx$ 5 mJ mol$^{-1}$ K$^{-2}$ [36], and $C_e = (\pi^2 k_B^2/3) T_0 D_0(\mu_0)$ (see Eq. (33)), yielding $D_0(\mu_0)/N = 3\gamma/\pi^2 k_B^2 \approx 2.1$ eV$^{-1}$ per Fe atom.

On the other hand, the absorbed pump energy per Fe atom can be calculated as $\Delta W_{abs}/N = F_{abs} v/d$, where $F_{abs}$ is the absorbed pump fluence, $d = 4$ nm is the Fe film thickness, and $v = V/N = 0.012$ nm$^3$ is the volume per Fe atom. Combining these considerations with Eqs. (38) and (39) allows us to determine $C$ and yields the change in the electron occupation number directly after pump absorption,

$$\Delta n(\epsilon) = \frac{F_{abs} v/d}{D_0(\mu_0)/N} \frac{n_0(\epsilon - \hbar \omega_p) - 2n_0(\epsilon) + n_0(\epsilon + \hbar \omega_p)}{\int d\epsilon \, (\epsilon - \mu_0)[n_0(\epsilon - \hbar \omega_p) - 2n_0(\epsilon) + n_0(\epsilon + \hbar \omega_p)]}. \tag{40}$$

Figure 1(b) of the main text shows calculations of $\Delta n(\epsilon)$ for optical (3.1 eV) and THz (4.1 meV) photons for an absorbed fluence of $F_{abs} = 0.01$ mJ cm$^{-2}$. This value is slightly smaller than the THz absorbed fluence in the experiment (0.017 mJ cm$^{-2}$) since the THz pulse duration is comparable to the electron-phonon thermalization time. Fitting the resulting $\Delta n(\epsilon)$ by a Fermi function (temperature $T_0 + \Delta T_e$ and chemical potential $\mu_0$) minus $n_0(\epsilon)$ yields a change of $\Delta T_e = 140$ K in the electron temperature [see Fig. 1(b)]. For comparison, using the electron specific heat [44], we obtain $\Delta T_e = \Delta W_{abs}/C_e = 120$ K, which is in good agreement with the fit approach.

**Acknowledgments**

We acknowledge funding by the German Science Foundation through the collaborative research center SFB TRR 227 "Ultrafast spin dynamics" (projects A01, A05, B02 and B03) and the European Union through the ERC H2020 CoG project TERAMAG/Grant No. 681917.